\documentstyle[12pt]{article}
\begin{document}
\rightline{\bf imsc/93-44}
\rightline{\bf SFU.HEP.109/1993}
\begin{center}
{\bf HARMONIC MAPS AND SELF-DUAL EQUATIONS FOR IMMERSED SURFACES}
\end{center}
\vspace{4cm}

\begin{center}
R.Parthasarathy \\
The Institute of Mathematical Sciences \\
Madras 600 113, India \\
and \\
K.S.Viswanathan \\
Department of Physics \\
Simon Fraser University \\
Burnaby. B.C, Canada V5A 1S6
\end{center}

\newpage

\noindent {\it {Abstract}}

   The immersion of the string world sheet, regarded as a Riemann surface, in
$R^3$ and $R^4$ is described by the generalized Gauss map. When the Gauss map
is harmonic or equivalently for surfaces of constant mean  curvature, we
obtain Hitchin's self-dual equations, by using $SO(3)$ and $SO(4)$ gauge fields
constructed in our earlier studies. This complements our earlier result that
$h\surd g\ =\ 1$ surfaces exhibit Virasaro symmetry. The self-dual system so
obtained is compared with self-dual Chern-Simons system and a generalized
Liouville equation involving extrinsic geometry is obtained.

   The immersion in $R^n, \ n>4$ is described by the generalized Gauss map. It
is shown that when the Gauss map is harmonic, the mean curvature of the
immersed surface is constant. $SO(n)$ gauge fields are constructed from the
geometry of the surface and expressed in terms of the Gauss map. It is found
Hitchin's self- duality relations for the gauge group $SO(2)\times SO(n-2)$.

\newpage

\noindent {\bf I.INTRODUCTION}

\vspace{0.5cm}

    The study of Yang-Mills connections on Riemann surfaces is of importance in
string theory. The space of self-dual connections provides a model for
Teichm\"{u}ller space [1]. Clearly, the world sheet of a string is a
2-dimensional surface immersed in $R^n$ (For convenience we consider both the
world sheet and the target space as Euclidean). The immersion induces a metric
on the world sheet. The second fundamental form of the surface determines its
extrinsic geometry. We [2] have developed a formalism to study the dynamics of
the world sheet conformally immersed ( by conformal immersion it is meant that
the induced metric is in the conformal gauge ) in $R^3$ and $R^4$ using the
generalized Gauss map [3]. The extrinsic curvature action can be written as a
constrained Grassmannian $\sigma$-model action and the theory is asymptotically
free.

   Subsequently [4] we found a hidden Virasaro symmetry for surfaces of
constant scalar mean curvature density ($h\surd g = 1$). An action exhibiting
this symmetry has recently been constructed [5] and it is a WZNW action. The
quantum theory of this action has also been studied in [5]. It would be of
interest to know if surfaces of constant scalar mean curvature ($h$=constant)
exhibit novel properties, so that we get a better understanding of the many
facets of the string world sheet. For such surfaces, it is known that the
Gauss map from the world sheet $M$ (regarded as a Riemann surface) into the
Grassmannian $G_{2,n} \simeq SO(n)/(SO(2)\times SO(n-2))$ is harmonic for
$n$ = 3,4. It is the purpose of this paper to show that there exists a
non-Abelian self-dual system for such surfaces.

   It is instructive to compare our results with those of Hitchin [1] and
Donaldson [6]. Hitchin [1] has obtained two dimensional self-dual
Yang-Mills-Higgs system by dimensional reduction of the Euclidean 4-dimensional
self-dual $SU(2)$ Yang-Mills fields. By dimensional reduction here, one means
that the fields are independent of $x_3$ and $x_4$. The complex Higgs field is
identified with $\mu = 3$ and $4$ component of the Yang-Mills field.
Donaldson  considers harmonic map from a Riemann surface $M$ into
$H^{3}$,an hyperbolic 3-space of constant negative curvature. The
harmonic map extremizes the `energy functional'.   By showing an
existence of a flat $PSL(2,C)$ connection and using the harmonic map
equation ( the Euler-Lagrange equation ), Hitchin's equations were obtained.

   In this paper, we first consider the Gauss map of a 2-dimensional surface
into the Grassmannian $G_{2,n}\simeq SO(n)/(SO(2)\times SO(n-2))$ for $n=3,4$.
We [4] have previously constructed $SO(n)$ connections on the string world
sheet. We project these onto $SO(2)\times SO(n-2)$ and the coset. The
projection onto $SO(2)\times SO(n-2)$ is identified as the gauge field (see
sec.III) and that on the coset as the Higg's field. Using the Euler-Lagrange
equations (harmonic map equations) for the surface, we show that this system
is self-dual. In this analysis, the harmonic map has a geometrical
interpretation. For immersion in $R^3$ and $R^4$, the Gauss map is harmonic if
the mean curvature scalar $h$ is constant [7]. We next take up immersion in
$R^n$ for $n>4$ and show explicitly that when the Gauss map is harmonic, the
mean scalar curvature of the surface is constant. The $SO(n)$ gauge fields are
constructed from the geometry of the surface. Using their projections onto the
subgroup $SO(2)\times SO(n-2)$ and its complement in $G_{2,n}$, the self-dual
system of Hitchin is obtained when and only when the Gauss map is harmonic.
Thus our main result is:

\noindent {\bf Theorem.1}

  Let $M$ be an oriented surface immersed in $R^{n}$ and let $M_{0}$ be
  the Riemann surface obtained by the induced conformal structure on $M$.
  Let ${\cal{G}} : M_{0} \rightarrow G_{2,n}$ be the Gauss map. Let $A$ be the
  flat $SO(n)$ connection induced on $M_{0}$ defined by the adapted frame
  of the tangents and (n-2) normals to $M$. The projection of $A$ expressed
  in terms of the coordinates of the quadric $Q_{n-2}$ taken as the model
  for $G_{2,n}$, onto $SO(n-2)\times SO(2)$ and its complement in $G_{2,n}$
  satisfy Hitchin's self dual system of equations when the Gauss map is
  harmonic.

\vspace{0.5cm}

   Recently, Dunne, Jackiw, Pi and Trugenberger [8] made a systematic analysis
of the Yang-Mills non-linear Schr\"{o}dinger equation and demonstrated
self-dual Chern-Simons equations for  static configurations. Here the matter
density is in the adjoint representation. By choosing the Chern-Simons gauge
field in the commuting set of the Cartan subalgebra and $\Psi$ ($\rho =
-i[\Psi, {\Psi}^{\dagger}]$) in terms of the ladder operators with positive
roots, they [8] and Dunne [9] obtain Toda equations. We obtain similar results
for the Gauss map.

   In this way, we find that the (extrinsic) geometry of the string world sheet
is closely related to the self-dual system of Hitchin [1], Donaldson [6] and to
the static configuration of (2+1) self-dual non-Abelian Chern-Simons theory at
the classical level.

\vspace{1.5cm}

\noindent {\bf II.PRELIMINARIES}

\vspace{0.5cm}

   Consider a 2-dimensional (Euclidean) string world sheet $M$ immersed
in $R^{n}$. Let $M_{0}$ be the Riemann surface obtained by the induced
conformal structure on $M$. The induced metric is $g_{\alpha \beta}\ =\
{\partial}_{\alpha}X^{\mu}.{\partial}_{\beta}X^{\mu}$,
with $X^{\mu}({\xi}_1,{\xi}_2)$ as immersion coordinates ($\mu =$ 1,2,....n)
and ${\xi}_1,{\xi}_2$ as local isothermal coordinates on the surface. The
Gauss-Codazzi equations introduce the second fundamental form
$H^i_{\alpha\beta}$, $i=$ 1,2,...(n-2). Locally on the surface, we have two
tangents and (n-2) normals. The Gauss map is,
\begin{eqnarray}
{\cal{G}} : M_{0}\rightarrow G_{2,n}&\simeq & SO(n)/(SO(2)\times SO(n-2)).
\end{eqnarray}
$G_{2,n}$ admits a complex structure. It is convenient to regard $G_{2,n}$ as a
quadric in $CP^{n-1}$ defined by $\sum_{i=1}^n Z^2_i\ =\ 0$, where $Z_i$ are
the homogeneous coordinates on $CP^{n-1}$. A local tangent 2-plane to $M$ is an
element of $G_{2,n}$, or equivalently a point in $Q_{n-2}$. Then [3] we have,
\begin{eqnarray}
{\partial}_z X^{\mu} &=& \psi {\Phi}^{\mu},
\end{eqnarray}
where $z={\xi}_1 + i{\xi}_2$, $\bar{z} = {\xi}_1 - i{\xi}_2$, ${\Phi}^{\mu}\in
Q_{n-2}$,${\Phi }^{\mu} {\Phi }^{\mu}=0$ and $\psi $ is a complex function of
$z$ and $\bar{z}$ which can be determined in terms of the geometrical
properties of the surface. As not every element of $G_{2,n}$ is a tangent
plane to $M$, the Gauss map (2) has to satisfy (n-2) conditions of
integrability [3]. These were explicitly obtained in Ref.3 for immersion in
$R^3$ and $R^4$ and by us [11] for $R^n\  (n>4)$. We first consider immersion
in
$R^3$ and $R^4$.

   For immersion in $R^3$, ${\Phi}^{\mu}$ is parametrized as,
\begin{eqnarray}
{\Phi}^{\mu} &=& \left[ 1-f^2, i(1+f^2), 2f \right],
\end{eqnarray}
where $f$ is complex. The Gauss map integrability condition is,
\begin{eqnarray}
Im \left[ \frac{f_{z\bar{z}}}{f_{\bar{z}}} - \frac{2\bar{f}f_z}{1+{\mid
f\mid}^2} \right]_{\bar{z}} &=& 0.
\end{eqnarray}
The mean curvature  $h (=N^{\mu} H^{\mu \alpha}_{\alpha})$ is given by,
\begin{eqnarray}
(\ell n h)_z &=& \frac{f_{z\bar{z}}}{f_{\bar{z}}} - \frac{2\bar{f}f_z}{1+
{\mid f \mid}^2},
\end{eqnarray}
which is known as the Kenmotsu equation [10]. The normal $N^{\mu}$ to the
surface can be expressed in terms of $f$ as,
\begin{eqnarray}
N^{\mu} &=& \frac{1}{1+{\mid f\mid}^2}\left[ f+\bar{f}, -i(f-\bar{f}),{\mid f
\mid}^2 - 1\right].
\end{eqnarray}
The energy integral of the surface is,
\begin{eqnarray}
S&=&\int \frac{{\mid f_{\bar{z}}\mid }^2 + {\mid f_z \mid}^2}{(1+{\mid f
\mid}^2)^2} dz d\bar{z},
\end{eqnarray}
which is also the extrinsic curvature action $\int \surd g {\mid H\mid}^2$. The
Euler-Lagrange equations are,
\begin{eqnarray}
L(f) &=& f_{z\bar{z}} - \frac{2\bar{f}f_zf_{\bar{z}}} {1+{\mid f\mid}^2} = 0.
\end{eqnarray}

   {\it {The Gauss map is said to be harmonic if $f$ satisfies the Euler-
Lagrange equations and it then follows from (5) that $h$ is constant [7]}}.

   For immersion in $R^4$, we have $G_{2,4} \simeq SO(4)/(SO(2)\times
SO(2))\simeq CP^1 \times CP^1$ and so ${\Phi}^{\mu}$ is parametrized in terms
of the two $CP^1$ fields, $f_1$ and $f_2$ as,
\begin{eqnarray}
{\Phi}^{\mu} &=& \left[ 1+f_1f_2, i(1-f_1f_2), f_1 - f_2, -i(f_1 + f_2)\right].
\end{eqnarray}
The Gauss map integrability conditions are,
\begin{eqnarray}
Im\left[ \sum^{2}_{i=1} \frac{f_{iz\bar{z}}}{f_{i\bar{z}}} - \frac{2{\bar{f}}_i
f_{iz}}{1+{\mid f_i \mid}^2}\right]_{\bar{z}} &=& 0, \nonumber \\
\mid F_1 \mid &=& \mid F_2 \mid,
\end{eqnarray}
where $F_i = \frac{f_{i\bar{z}}}{1+{\mid f_i \mid}^2}$. There are two normals
$N^{\mu}_1\ ,\ N^{\mu}_2$ to the surface which can be written in terms of
$f_1$ and $f_2$ as,
\begin{eqnarray}
D &=& \left( (1+{\mid f_1 \mid}^2)(1+{\mid f_2 \mid}^2)\right)^{\frac{1}{2}},
\nonumber \\
A^{\mu} &=& \left[f_2-{\bar{f}}_1, -i(f_2+{\bar{f}}_1), 1+{\bar{f}}_1f_2,
-i(1-{\bar{f}}_1f_2)\right], \nonumber \\
N^{\mu}_1 &=& \frac{1}{2} (A^{\mu} + {\bar{A}}^{\mu})/D, \nonumber \\
N^{\mu}_2 &=& \frac{1}{2i}(A^{\mu} - {\bar{A}}^{\mu})/D.
\end{eqnarray}
Then the components of $H^{\mu \alpha}_{\alpha}$ along $N^{\mu}_1$ and
$N^{\mu}_2$ are given by,
\begin{eqnarray}
h_1 &=& \frac{F_1 - F_2}{2\bar{\psi}D} \nonumber \\
h_2 &=& \frac{i(F_1 + F_2)}{2\bar{\psi}D},
\end{eqnarray}
and the mean curvature  $h^2\ =\ (h^2_{1} + h^2_{2})$  satisfies
the equation [2],
\begin{eqnarray}
2(\ell n h)_z &=& \sum^{2}_{i=1} \left[ \frac{f_{iz\bar{z}}}{f_{i\bar{z}}} -
\frac{2\bar{f}_if_{iz}}{1+{\mid f_i \mid}^2}\right].
\end{eqnarray}
The energy integral of the surface is,
\begin{eqnarray}
S &=& \int \sum^{2}_{i=1} {\mid F_i\mid }^2 + {\mid \hat{F}_i \mid}^2,
\end{eqnarray}
where $\hat{F}_i = \frac{f_{iz}}{1+{\mid f_i \mid}^2}$. The Euler-Lagrange
equations are obtained as,
\begin{eqnarray}
L(f_1) = 0 &;& L(f_2) = 0.
\end{eqnarray}
The Gauss map is harmonic if $f_1$ and $f_2$ satisfy (15) and by (13)
harmonicity of the Gauss map from $M$ into $G_{2,4}$ implies that the immersed
surface has constant $h$ for immersion in $R^4$. It is to be noted that $f_1$
and $f_2$ should also satisfy the second requirement in (10) for them to
describe the Gauss map.

   In Ref.4, we have considered tangents to $M$ as ${\hat{e}}_1\ =\ \frac{1}{
\surd 2 \mid \Phi \mid}({\Phi}^{\mu} + {\bar{\Phi}}^{\mu})$ and ${\hat{e}_2}\
 =\ \frac{1}{\surd 2 i\mid \Phi \mid}({\Phi}^{\mu} - {\bar{\Phi}}^{\mu})$ along
with the (n-2) normals. Then, the local orthonormal frame $\left( {\hat{e}}_1,
{\hat{e}}_2, N^{\mu}_i \right)$ satisfies,
\begin{eqnarray}
{\partial}_z {\hat{e}}_i &=& (A_z)_{ij} {\hat{e}}_j;\ \ \ \ i,j\ =\ 1\  to\  n,
\end{eqnarray}
where ${\hat{e}}_i\ =\ N^{\mu}_i$, for $i\ =\ 3$ to $n$ and $(A_z)_{ij}$ is an
antisymmetric $n\times n$ matrix. A similar equation for the $\bar{z}$
derivative defines $(A_{\bar{z}})_{ij}$. $A_z$ and $A_{\bar{z}}$ are easily
seen to transform as $SO(n,C)$ gauge fields under local $SO(n)$ transformations
of $({\hat{e}}_1, {\hat{e}}_2, N^{\mu}_i)$ which follows from (16) [4]. These
non-Abelian gauge fields are constructed from the geometrical properties of the
surface alone and so they are characteristics of the world sheet. Using
equations (3) and (6), it is easily verified that $A_z$ for immersion in $R^3$
is given by,

\begin{eqnarray}
A_z&=&\frac{1}{1+{\mid f\mid}^2}\left[ \begin{array}{lcr}
0 & -i(f{\bar{f}}_z - \bar{f}f_z) & -(f_z + {\bar{f}}_z) \\
i(f{\bar{f}}_z - \bar{f}f_z) & 0 & i(f_z - {\bar{f}}_z) \\
f_z + {\bar{f}}_z & -i(f_z - {\bar{f}}_z) & 0
\end{array} \right]
\end{eqnarray}

Similarly, using equations (9) and (10), and denoting $d_i=1+{\mid f_i\mid}^2$,
$m_i=f_i{\bar{f}}_{iz}-{\bar{f}}_if_{iz}$, $p_i=f_{iz} +{\bar{f}}_{iz}$, and
$q_i=f_{iz}-{\bar{f}}_{iz}$, $A_z$ for immersion in $R^4$ is,

\begin{eqnarray}
A_z&=&\frac{1}{2}\left[ \begin{array}{lccr}
0 &-i(\frac{m_1}{d_1}+\frac{m_2}{d_2})&\frac{p_1}{d_1}-\frac{p_2}{d_2}&i(\frac
{q_1}{d_1} +\frac{q_2}{d_2} \\
i(\frac{m_1}{d_1}+\frac{m_2}{d_2})&0&-i(\frac{q_1}{d_1}-\frac{q_2}{d_2})&\frac
{p_1}{d_1}+\frac{p_2}{d_2} \\
-(\frac{p_1}{d_1}-\frac{p_2}{d_2})&i(\frac{q_1}{d_1}-\frac{q_2}{d_2})&0&i(\frac
{m_1}{d_1}-\frac{m_2}{d_2}) \\
-i(\frac{q_1}{d_1}+\frac{q_2}{d_2})&-(\frac{p_1}{d_1}+\frac{p_2}{d_2})&-i(\frac
{m_1}{d_1}-\frac{m_2}{d_2})&0
\end{array} \right]
\end{eqnarray}

   The gauge field $A_{\bar{z}}$ can be obtained by replacing $z$ derivatives
by $\bar{z}$ derivatives and it is seen that $(A_z)^{\dagger}\ =\ -A_{\bar{z}}$
. Further, from (16) it is easily verified that the gauge fields satisfy,
\begin{eqnarray}
{\partial}_{\bar{z}}A_z - {\partial}_zA_{\bar{z}} + [A_z,A_{\bar{z}}]&=&0.
\end{eqnarray}

\noindent {\bf III.HARMONIC MAP AND SELF-DUAL SYSTEM}

\vspace{0.5cm}

   We now project the gauge fields constructed in the previous section onto
$SO(2)\times SO(n-2)$ and its orthogonal complement in $G_{2,n}$ for $n \ =$
3 and 4. The general procedure is briefly outlined here. (For details see
Ref.13). Consider a $G/H$ sigma model on a two dimensional Riemann surface.
Denote the generators of the Lie algebra $L_G$ of $G$ by $L(\tilde{\sigma})$,
$\tilde{\sigma}=$ 1,2,...[G] and those of $L_H$ of $H$ by $L(\bar{\sigma})$;
$\bar{\sigma}$= 1,2,....[H]; [H]$<$[G]. The remaining generators of $L_G$ will
be denoted by $L(\sigma)$. Consider a local gauge group associated with $G$. We
have,
\begin{eqnarray}
M\ \ni \ (z,\bar{z})&\stackrel{g}{\rightarrow}& g(z,\bar{z}) \in G,
\end{eqnarray}
and introduce,
\begin{eqnarray}
{\omega}_{\alpha}(g) &=& g^{\dagger}{\partial}_{\alpha}g.
\end{eqnarray}
The field strength associated with ${\omega}_{\alpha}(g)$ is zero. In fact,
${\omega}_{\alpha}(g)$ is same as $-A_z$ and $-A_{\bar{z}}$ and  is equivalent
to (16) with $g(z,\bar{z})$ as the $n\times n$ matrix formed by the two tangent
vectors ${\hat{e}}_1$ and ${\hat{e}}_2$ and the $(n-2)$ normals $N^{\mu}_i$.
Under a local gauge transformation generated by $u(z,\bar{z})\in H$, we have,
\begin{eqnarray}
g(z,\bar{z})&\rightarrow & g(z,\bar{z})u(z,\bar{z}), \nonumber \\
{\omega}_{\alpha}(g)\rightarrow
{\omega}_{\alpha}(gu)&=&u^{\dagger}{\omega}_{\alpha}(g)u +
u^{\dagger}{\partial}_ {\alpha}u.
\end{eqnarray}
Thus $-A_z$ and $-A_{\bar{z}}$ transform as gauge fields under $SO(2)\times
SO(n-2)$ gauge transformation. The projection of ${\omega}_{\alpha}(g)$ onto
$L_H$ and its orthogonal complement are,
\begin{eqnarray}
a_{\alpha}(g)&=&L(\bar{\sigma}) tr (L(\bar{\sigma}){\omega}_{\alpha}(g)),
\nonumber \\
b_{\alpha}(g)&=&L(\sigma) tr (L(\sigma){\omega}_{\alpha}(g)),
\end{eqnarray}
and it is straightforward to verify that under (22),
\begin{eqnarray}
a_{\alpha}(g)\rightarrow a_{\alpha}(gu) &=& u^{\dagger}a_{\alpha}u +
u^{\dagger}{\partial}_{\alpha}u, \nonumber \\
b_{\alpha}(g)\rightarrow b_{\alpha}(gu) &=& u^{\dagger}{\partial}_{\alpha}u.
\end{eqnarray}
So, $a_{\alpha}(g)$ transforms as a gauge field under local gauge
transformations  belonging to $H$ and $b_{\alpha}(g)$ transforms homogeneously.

   Now we consider immersion in $R^3$. The $SO(3)$ gauge fields $A_z$ and
$A_{\bar{z}}$ in (17) are projected onto $SO(2)$ and its orthogonal complement
in $G_{2,3}$. Denoting the anti-Hermitian generators of $SO(3)$ as
$T_1,T_2,T_3$, $[T_1,T_2]\ =\ T_3$, et .cyc, we have,
\begin{eqnarray}
a_z &=& \frac{1}{2} T_3 tr (T_3 A_z), \nonumber \\
b_z &=& \frac{1}{2} T_1 tr (T_1 A_z) + \frac{1}{2} T_2 tr (T_2 A_z).
\end{eqnarray}
The gauge group in (22) is $U(1)$. Similar projections for $A_{\bar{z}}$ are
made. It can be verified that $a_z\ +\ b_z\ =\ -A_z\ =\ g^{\dagger}{\partial}_z
g$. Once we have a flat connection which can be decomposed as $a_z$ and $b_z$ (
and similarly for $A_{\bar{z}}$), we have,
\begin{eqnarray}
{\partial}_z a_{\bar{z}} - {\partial}_{\bar{z}} a_z + [a_z, a_{\bar{z}}] +
[b_z, b_{\bar{z}}] &=& 0, \nonumber \\
{\partial}_{\bar{z}} b_z + [a_{\bar{z}}, b_z] &=& {\partial}_z b_{\bar{z}} +
[a_z, b_{\bar{z}}],
\end{eqnarray}
where we have made use of the group structure underlying (25), namely; the
first equation in (26) is in the Cartan subalgebra while the second in $T_1$
and $T_2$ directions: hence both must separately vanish. The second equation in
(26) gives the self-dual property if each side vanishes. This we shall prove by
using the equations of motion (15), namely for harmonic Gauss map. Explicitly,
from (17), we find,
\begin{eqnarray}
a_z &=& \frac{1}{1+{\mid f\mid }^2}\left[ \begin{array}{lcr}
0 & i(f{\bar{f}}_z-\bar{f}f_z) & 0 \\
-i(f{\bar{f}}_z-\bar{f}f_z) & 0 & 0 \\
0 & 0 & 0
\end{array} \right],
\end{eqnarray}
and,
\begin{eqnarray}
b_z &=& \frac{1}{1+{\mid f\mid}^2}\left[ \begin{array}{lcr}
0 & 0 & f_z + {\bar{f}}_z \\
0 & 0 & -i(f_z - {\bar{f}}_z) \\
-(f_z+{\bar{f}}_z) & i(f_z - {\bar{f}}_z) & 0
\end{array} \right].
\end{eqnarray}
$a_{\bar{z}}\ =\ - {a_z}^{\dagger}$ ; $b_{\bar{z}}\ =\ -{b_z}^{\dagger}$. Then
we find that,
\begin{eqnarray}
{\partial}_{\bar{z}}b_z + [a_{\bar{z}}, b_z] &=& 0,
\end{eqnarray}
if and only if $L(f)\ =\ 0$. Thus for surfaces of constant $h$, i-e for
harmonicGauss maps, we find that,
\begin{eqnarray}
{\partial}_z a_{\bar{z}} - {\partial}_{\bar{z}}a_z + [a_z, a_{\bar{z}}] + [b_z,
b_{\bar{z}}] &=& 0, \nonumber \\
{\partial}_{\bar{z}}b_z + [a_{\bar{z}}, b_z] &=& 0, \nonumber \\
{\partial}_zb_{\bar{z}} + [a_z, b_{\bar{z}}] &=& 0.
\end{eqnarray}
Further $a_z$ transforms as an $SO(2)$ gauge field while $b_z$ transforms
homogeneously. We have thus obtained Hitchin's self-dual Yang-Mills-Higgs
system for harmonic maps in $G_{2,3}$. Equations (30) are also equivalent to
static self-dual Chern-Simons system if we set the matter density $\rho \ =\
-i[b_z, {b_z}^{\dagger}]$ which lies in the Cartan subalgebra of $SO(3)$.
Stated differently, we have thus shown that harmonic Gauss maps of
immersed surfaces in $R^{3}$ represented by $f(z)$ in Eqn.(3) satisfies
the self-dual equations of Hitchin.

   Next consider surfaces in $R^4$. Here the Gauss map maps $M_{0}$ into \\
$G_{2,4}\simeq SO(4)/(SO(2)\times SO(2))$. We choose $T_1$ to $T_6$ as
generators of $SO(4)$ such that $[T_1, T_2]\ =\ T_3$, et.cyc; $[T_4,T_5]\ =\
T_6$, et.cyc; and $[T_i,T_j]\ =\ 0$ for $i\ =\ 1,2,3$; $j\ =\ 4,5,6$. The
explicit form of $A_z$ has been given in Eqn.(18) and the projection of $A_z$
and $A_{\bar{z}}$ onto $SO(2)\times SO(2)$ and its complement in $G_{2,4}$ are,
\begin{eqnarray}
a_z&=& T_3 tr (T_3 A_z) + T_6 tr (T_6 A_z), \nonumber \\
b_z &=& T_1 tr (T_1A_z) + T_2 tr (T_2 A_z) + T_4 tr (T_4 A_z) + T_5 tr (T_5
A_z).
\end{eqnarray}
Equations similar to (26) readily follow from (18). The explicit forms of $a_z$
and $b_z$ are not displayed as the procedure is straightforward. The self-dual
property can be verified for harmonic maps by computing each side of the second
equation in (26) for this case, immersion in $R^4$. Introducing,
\begin{eqnarray}
{\cal L}(f_i) &=& \frac{(L(f_i) + \bar{L}(f_i))}{1+{\mid f_i\mid}^2}, \nonumber
\\
{\cal L}'(f_i)&=& \frac{(L(f_i) - \bar{L}(f_i))}{1+{\mid f_i\mid}^2}, \nonumber
\\
{\cal S} &=& {\cal L}(f_1) + {\cal L}(f_2), \nonumber \\
{\cal D} &=& {\cal L}(f_1) - {\cal L}(f_2), \nonumber \\
{\cal S}' &=& {\cal L}'(f_1) + {\cal L}'(f_2), \nonumber \\
{\cal D}' &=& {\cal L}'(f_1) - {\cal L}'(f_2),
\end{eqnarray}
for $i\ =\ 1,2$ and where $L(f)$ is defined in (8), we find,
\begin{eqnarray}
{\partial}_{\bar{z}}b_z+[a_{\bar{z}},b_z]&=&\frac{1}{2}\times
   \left[ \begin{array}{lccr}
0 & 0 & -{\cal D} & -i{\cal S}' \\
0 & 0 & i{\cal D}' &-{\cal S} \\
{\cal D} & -i{\cal D}' &0 &0 \\
i{\cal S}' & {\cal S}& 0 &0
\end{array} \right].
\end{eqnarray}
It can be seen that when the Euler-Lagrange equations of motion are satisfied,
$L(f_i) = 0$ for $i=1,2$ it follows that ${\partial}_{\bar{z}}b_z\ =\
[a_{\bar{z}}, b_z]=0$, which is the self-dual equation. It is pertinent to
note that $b_z$ transforms homogeneously under the local $SO(2)\times SO(2)$
gauge transformation. The field $a_z$ which is in the Cartan subalgebra
$SO(2)\times SO(2)$ transforms as a gauge field. It is important to
reiterate that $a_z$ and $b_z$ are embedded in $SO(4)$. In this way we realize
Hitchin's equations for two dimensional gauge fields (constructed from the
surface  itself) on the world sheet immersed in $R^4$, when the Gauss map is
harmonic. Conversely, explicit solutions to the self-dual equations for the
gauge group studied here are given by (27) and (28) for $R^3$ and (18) and (31)
for $R^4$, where the complex functions $f_1$ and $f_2$ satisfy the equations,
$L(f_1)=0$ and $L(f_2)=0$. Since we have seen that harmonicity of the
Gauss map implies that the surfaces have constant mean curvature, it can
be concluded that such surfaces possess the properties of a self dual
2-dimensional field theory.

   The basic equations (30) are very similar to the self-dual Chern-Simons
system considered in Ref.8. Fujii [14] examined the relationship between Toda
systems and the Grassmannian $\sigma$-models. We now extend our cosiderations
for $R^3$. In anology with [8], the matter density $\rho \ =\ {\rho}_3 T_3$ is,
\begin{eqnarray}
{\rho}_3 &=& \frac{2(f_z{\bar{f}}_{\bar{z}} - f_{\bar{z}}{\bar{f}}_z)}{(1+{\mid
f \mid}^2)^2}.
\end{eqnarray}
We recall the following relations for Gauss map in $R^3$. With $\surd g\ =\
exp( \phi)$, $\hat{F}\ =\ \frac{f_z}{1+{\mid f\mid}^2}$, we have,
\begin{eqnarray}
{\mid F\mid}^2&=& \frac{h}{2}H_{z\bar{z}} = \frac{h^2}{2}\exp(\phi), \nonumber
\\
{\mid \hat{F}\mid}^2&=& \frac{h}{2}
\frac{H_{zz}H_{\bar{z}\bar{z}}}{H_{z\bar{z}}}=\frac{1}{2}{\mid
H_{zz}\mid}^2\exp(-\phi), \nonumber \\
H_{z\bar{z}}&=&h\surd g = h\exp(\phi), \nonumber \\
{\rho}_3 &=& 2({\mid \hat{F}\mid}^2 - {\mid F\mid}^2),
\end{eqnarray}
where $H_{\alpha \beta}$ is the second fundamental form. It can be verified
by using the equation of motion (8) and the Gauss map relation, $(\ell
n\psi)_{\bar{z}}\ =\ -\frac{2\bar{f}f_{\bar{z}}}{(1+{\mid f\mid}^2)}$, that
${\partial}_z{\partial}_{\bar{z}}\ell n{\mid H_{zz}\mid}^2\ =\ 0$. The Gauss
curvature $R\ =\ -\exp(-\phi){\partial}_z{\partial}_{\bar{z}}\phi$. Writing the
Gauss curvature in terms of $H_{\alpha \beta}$, we have a modified Liouville
equation for extrinsic curvature as,
\begin{eqnarray}
{\partial}_z{\partial}_{\bar{z}}\phi &=& -2h^2\exp(\phi) +
2\exp(-\phi)\exp({\phi}_E),
\end{eqnarray}
where ${\mid H_{zz}\mid}^2\ =\ \exp({\phi}_E)$ and
${\partial}_z{\partial}_{\bar{z}}{\phi}_E\ =\ 0$. When we consider $f$ to be
anti-holomorphic, then $\hat{F}\ =\ 0$ and (39) reduces to the Liouville
equation for $h\ =\ 1$,
\begin{eqnarray}
{\partial}_z{\partial}_{\bar{z}}\phi &=& -2\exp(\phi),
\end{eqnarray}
which is also the Toda equation for $SO(3)$. Hence Eqn.(36) may be viewed
as the generalized Liouville equation for immersed surfaces whose induced
Riemann surface has genus $g>1$. (See below)

   For immersion in $R^4$, when we consider $f_1$ and $f_2$ both
anti-holomorphic, we obtain,
\begin{eqnarray}
\rho &=& \frac{f_{1\bar{z}}{\bar{f}}_{1z}}{(1+{\mid f_1\mid}^2)^2} T_6 +
\frac{f_{2\bar{z}}{\bar{f}}_{2z}}{(1+{\mid f_2\mid}^2)^2} T_3.
\end{eqnarray}
The Cartan matrix for $SO(4)$ is $K_{\alpha\beta}\ =\
-2{\delta}_{\alpha\beta}$. Then we obtain,
\begin{eqnarray}
{\partial}_z{\partial}_{\bar{z}}\ell n {\rho}_6 &=& -2{\rho}_6, \nonumber \\
{\partial}_z{\partial}_{\bar{z}}\ell n {\rho}_3 &=& -2{\rho}_3,
\end{eqnarray}
where ${\rho}_6$ and ${\rho}_3$ are the coefficients of $T_6$ and $T_3$ in (42)
. When $f$ satisfies the Euler-Lagrange equation $L(f)\ =\ 0$, the immersed
surface has $h$ constant. If $f$ is anti-holomorphic, the Gauss map is a map
from $S^2$ into $G_{2,3}$. This can be seen by noting that,
\begin{eqnarray}
2\pi (1-g)&=&\int \frac{{\mid f_{\bar{z}}\mid}^2 - {\mid f_z\mid}^2}{(1+{\mid
f\mid}^2)^2} dz d\bar{z}, \nonumber \\
&=& \int \frac{{\mid f_{\bar{z}}\mid}^2}{(1+{\mid f\mid}^2)^2} dz d\bar{z},
\end{eqnarray}
where $g$ is the genus of the surface. Since the right hand side is positive
definite, it follows that $g=0$ unless $f=0$.

\vspace{1.5cm}
\noindent {\bf IV.HARMONIC MAPS IN $R^{n} (n>4)$}

\vspace{0.5cm}
   We now consider immersion of 2-dimensional surfaces in $R^n$, $n>4$. There
are two reasons for this consideration. First of all, the gauge field $a_z$
in the two cases considered (n=3 and 4) is Abelian embedding in $SO(3)$ and
$SO(4)$. This is similar to the choice in Ref.8 and 9. We would like to realize
the non-Abelian property of $a_z$ which occurs when $n>4$. Secondly, the result
that harmonic Gauss map implies constant mean curvature scalar, has been proved
for immersion in $R^3$ by Ruh and Vilms [7] and can be proved from our [2]
results for $h$ and the Euler-Lagrange equations for immersion in $R^4$. For
immersion in $R^n,\ n>4$, such a result has not yet been explicitly obtained to
the best of our knowledge. In this paper we prove this and use it to obtain the
self-duality equations for harmonic Gauss maps.

     We recall the essential details of the Gauss map of surfaces in $R^n$ from
our earlier paper [11] and from Hoffman and Osserman [10]. ${\Phi}^{\mu}$ of
$Q_{n-2}$ in (2) is parametrized in the following manner. Let $(z_1,z_2,\ .\ .\
.\ ., z_n)$ be the homogeneous coordinates of $CP^{n-1}$. The quadric
$Q_{n-2}\in CP^{n-1}$ is defined by,
\begin{eqnarray}
\sum_{k=1}^n {z_k}^2 &=& 0.
\end{eqnarray}
Let $H$ be the hyperplane in $CP^{n-1}$ defined by $H:(z_1 - iz_2)=0$. Then
$Q_{n-2}^* = Q_{n-2}\setminus \{H\}$ is biholomorphic to $C^{n-2}$ under the
correspondence [10],
\begin{eqnarray}
(z_1,.....,z_n)&=&\frac{z_1-iz_2}{2}\left[1-{{\zeta}_k}^2,i(1+{{\zeta}_k}^2),
2{\zeta}_1,.....,2{\zeta}_{n-2}\right],
\end{eqnarray}
where,
\begin{eqnarray}
{\zeta}_j &=& \frac{z_{j+2}}{z_1-iz_2},
\end{eqnarray}
for $j=1,2,....,n-2$. In (42) and in what follows we use the summation
convention that repeated indices are summed from $1$ to $n-2$, unless otherwise
stated. Conversely, given any $({\zeta}_1,...,{\zeta}_{n-2})\in C^{n-2}$, the
point,
\begin{eqnarray}
{\Phi}^{\mu}&=&\left[1-{{\zeta}_k}^2,i(1+{{\zeta}_k}^2),2{\zeta}_1,....,
2{\zeta}_{n-2}\right],
\end{eqnarray}
satisfies (41) and hence defines a point in the complex quadric $Q_{n-2}$. The
Fubini-Study metric on $CP^{n-1}$ induces a metric on $Q_{n-2}$ [11] which is
computed as,
\begin{eqnarray}
g_{ij}&=&\frac{4}{{\mid \Phi\mid}^2}{\delta}_{ij} + \frac{16}{{\mid
\Phi\mid}^4}
[{\zeta}_i{\bar{\zeta}}_j-{\zeta}_j{\bar{\zeta}}_i+2{\zeta}_i{\bar{\zeta}}_j
{\mid
{\zeta}_k\mid}^2-{\zeta}_i{\zeta}_j{{\bar{\zeta}}_k}^2-{\bar{\zeta}}_i{\bar
{\zeta}}_j{{\zeta}_k}^2],
\end{eqnarray}
where,
\begin{eqnarray}
{\mid \Phi\mid}^2&=&2+4{\zeta}_k{\bar{\zeta}}_k+2{{\zeta}_k}^2{{\bar{\zeta}}_m
}^2.
\end{eqnarray}
We [11] found it convenient to introduce an n-vector,
\begin{eqnarray}
{A}^{\mu}_k&=& -[{\bar{\zeta}}_k+{\zeta}_k{{\bar{\zeta}}_m}^2]{\Phi}^{\mu} +
\frac{{\mid \Phi\mid}^2}{2}  {v}^{\mu}_k,
\end{eqnarray}
for $k=1,2...(n-2)$ and,
\begin{eqnarray}
{v}^{\mu}_k&=&(-{\zeta}_k, i{\zeta}_k,0,0,0,..1_k,0,..),
\end{eqnarray}
where $1_k$ stands for 1 in the $(k+2)$th position. The algebraic properties of
${A}^{\mu}_k$ and ${a}^{\mu}_k$ have been established in [11]. The $(n-2)$ real
normals to the surface have been obtained as,
\begin{eqnarray}
{N}^{\mu}_i &=& \frac{4}{{\mid \Phi\mid}^4} (O^T)_{ij} {A}^{\mu}_j,
\end{eqnarray}
where the $(n-2)\times (n-2)$ matrix $O$ has been defined in [11].

   The $(n-2)$ complex functions ${\zeta}_i(z,\bar{z})$ where $z,\bar{z}$ are
the isothermal coordinates on $M$ have been shown to satisfy $(n-2)$ conditions
so that they can represent the Gauss map [11]. The mean curvature scalar $h$
of the surface has been shown to be related to Gauss map by,
\begin{eqnarray}
(\ell n h)_z&=&
\frac{\sum^{n-2}_{j=1}{\zeta}_{j\bar{z}}{\zeta}_{jz\bar{z}}}{\sum^{n-2}_{j=1}
({\zeta}_{j\bar{z}})^2}-\frac{4}{{\mid
\Phi\mid}^2}\sum^{n-2}_{j=1}{\zeta}_{jz}[{\bar{\zeta}}_j+{\zeta}_j\sum^{n-2}_
{k=1}{{\bar{\zeta}}_k}^2],
\end{eqnarray}
which is the generalization of the Kenmotsu equation to immersion in $R^n$.

\vspace{0.5cm}

\noindent {\bf Theorem.2}

   Let $M$ and $M_{0}$ be as defined in Theorem 1. Then, if the Gauss map
${\cal {G}}:M\rightarrow G_{2,n}$
is harmonic,  the mean curvature scalar $h$ of $M$ is constant.

\vspace{0.5cm}

\noindent {\it Proof}

   The Gauss map is said to be harmonic if the (n-2) complex functions
${\zeta}_i$ satisfy the Euler-Lagrange equations of the `energy integral'[11],
\begin{eqnarray}
{\cal {E}} &=& \int g_{ij} {\zeta}_{i\bar{z}}{\bar{\zeta}}_{jz}.
\end{eqnarray}
The above `energy integral' is also the action for the extrinsic curvature of
the surface $M$, namely, $\int \surd g {\mid H\mid}^2$. The Euler-Lagrange
equations that follow from the extremum of ${\cal{E}}$ are given by,
\begin{eqnarray}
{\zeta}_{kz\bar{z}}&=& -\frac{4}{{\mid \Phi\mid}^2} \sum^{n-2}_{i=1}
{\zeta}_{i\bar{z}} {\zeta}_{iz} ({\bar{\zeta}}_k +
{\zeta}_k\sum^{n-2}_{m=1}{{\bar{\zeta}}_m}^2) \nonumber \\
& + & \frac{4}{{\mid \Phi\mid}^2}\sum^{n-2}_{i=1}({\bar{\zeta}}_i +
{\zeta}_i\sum^{n-2}_{m=1}{{\bar{\zeta}}_m}^2) [{\zeta}_{k\bar{z}}{\zeta}_{iz} +
{\zeta}_{kz}{\zeta}_{i\bar{z}}].
\end{eqnarray}
Upon using the above Euler-Lagrange equations in the expression for $(\ell n
h)_z$ in (51) it follows that the mean curvature scalar $h$ of the surface $M$
is constant. Q.E.D

\vspace{0.5cm}

   We now construct $SO(n)$ gauge fields on the surface $M$. The defining
equation for them is (16) with ${\Phi}^{\mu}$ given by (44) and the $(n-2)$
normals by (49). The various components of $A_z$ are given below.
\begin{eqnarray}
A_z &=& \left[\begin{array}{lc|c}
0 & (A_z)_{12} & (A_z)_{1i}(i=3 \mbox { to } n) \\
(A_z)_{21} & 0 & (A_z)_{2i}(i=3 \mbox { to } n) \\
\hline
(A_z)_{31} & (A_z)_{32} & ........    \\
..&..&    (A_z)_{ij} \\
(A_z)_{n1} & (A_z)_{n2} &...
 \end{array}\right]
\end{eqnarray}
where,
\begin{eqnarray}
(A_z)_{12}&=& \frac{2}{i{\mid
\Phi\mid}^2}[({\zeta}_j+{{\bar{\zeta}}_j}{{\zeta}_i}^2){\bar{\zeta}}_{jz} -
({\bar{\zeta}}_j + {\zeta}_j{{\bar{\zeta}}_i}^2){\zeta}_{jz}] \nonumber \\
(A_z)_{1i}&=& \frac{2\surd 2}{{\mid \Phi\mid}^5} (O^T)_{ij}[{\mid
\Phi\mid}^2{\zeta}_{jz} -
4({\bar{\zeta}}_j+{\zeta}_j{{\bar{\zeta}}_m}^2)({\zeta}_k+{\bar{\zeta}}_k
{{\zeta}_q}^2){\bar{\zeta}}_{kz} \nonumber \\
&+& {\mid
\Phi\mid}^2(2{\zeta}_j{\bar{\zeta}}_k{\bar{\zeta}}_{kz}+{\bar{\zeta}}_{jz})]
\nonumber  \\
(A_z)_{2i}&=&\frac{2\surd 2}{i{\mid \Phi\mid}^5} (O^T)_{ij}[{\mid \Phi\mid}^2
{\zeta}_{jz} +
4({\bar{\zeta}}_j+{\zeta}_j{{\bar{\zeta}}_m}^2)({\zeta}_k+{\bar{\zeta}}_k
{{\zeta}_q}^2){\bar{\zeta}}_{kz} \nonumber \\
&-& {\mid \Phi\mid}^2(2{\zeta}_j{\bar{\zeta}}_k{\bar{\zeta}}_{kz} +
{\bar{\zeta}}_{jz})] \nonumber \\
(A_z)_{ij}&=&-\frac{1}{{\mid \Phi\mid}^2}{\partial}_z({\mid
\Phi\mid}^2){\delta}_{ij} + \frac{4}{{\mid
\Phi\mid}^4}(O^T)_{jk}{\partial}_zO_{ki}  \nonumber \\
&+&\frac{16}{{\mid
\Phi\mid}^6}({\bar{\zeta}}_k+{\zeta}_k{{\bar{\zeta}}_q}^2){\zeta}_{mz}[(O^T)
_{jk}(O^T)_{im}-(O^T)_{jm}(O^T)_{ik}],
\end{eqnarray}

$A_{\bar{z}}$ can be obtained by replacing the $z$-derivatives by $\bar{z}$
derivatives. They define the $SO(n)$ gauge fields on the surface $M$. We now
project them on to $SO(2)\times SO(n-2)$ and its orthogonal complement in
$G_{2,n} \simeq SO(n)/(SO(2)\times SO(n-2))$. Denoting these projections by
$a_z$ and $b_z$ respectively (and similarly for $a_{\bar{z}}$ and
$b_{\bar{z}}$), we have,
\begin{eqnarray}
a_z &=&- \left[ \begin{array}{lc|c}
0 & (A_z)_{12} & 0(13 \mbox { to } 1n) \\
(A_z)_{21} & 0 & 0(23 \mbox { to } 2n) \\
\hline
 0 & 0&   \\
.. & ..    &  (A_z)_{ij}   \\
 0 & 0&
\end{array}\right]
\end{eqnarray}
and,
\begin{eqnarray}
b_z &=& -\left[ \begin{array}{lc|c}
0 & 0 & (A_z)_{1i}(i=3 \mbox { to } n) \\
0 & 0 & (A_z)_{2i}(i=3 \mbox { to } n) \\
\hline
(A_z)_{31}&(A_z)_{32}&   \\
.. & ..&  0    \\
(A_z)_{n1}&(A_z)_{n2}&
\end{array}\right].
\end{eqnarray}

   From the general considerations described in (24), it follows that $a_z$
transforms as a gauge field under local $SO(2)\times SO(n-2)$ gauge
transformation while $b_z$ transforms homogeneously. It can be verified that
$a_z$ in (55) is indeed a non-Abelian gauge field when $n>4$. Further the
construction and the gauge group structure of $a_z$ are quite different from
those of Donaldson [6]. The gauge connection $a_z$ contains contributions from
both the tangent space and the normal frame to $M$ reflecting $SO(2)\times
SO(n-2)$ group structure. The orthogonal complement $b_z$ on the other hand
receives from interaction of tangents with the normals. As $b_z$ transforms
homogeneously under the local $SO(2)\times SO(n-2)$ gauge transformations, it
can be identified with the Higg's field of Hitchin. Realizing that $A_z\ =\
-(a_z \ +\ b_z)$ and using (19) we immediately obtain (26) exploiting the group
structure underlying (55) and (56). A similar feature has been used by
Donaldson [6]. In order to prove that second equation in (26) gives the self
duality, namely the vanishing of both the sides, we make use of the
Euler-Lagrange equation (52) or harmonic Gauss map. To prove the self-duality
equation, namely, ${\partial}_{\bar{z}}b_z\ +\ [a_{\bar{z}}\ ,\ b_z]\ =\ 0$,
when the Gauss map is harmonic, that is, when (52) is satisfied, we proceed as
below. We consider $(1i)$ component of the self-duality equation for $i\geq 3$
which follows from (55) and (56).
\begin{eqnarray}
({\partial}_{\bar{z}}b_z + [a_{\bar{z}},b_z])_{1i}&=&
{\partial}_{\bar{z}}(b_z)_{1i} + (a_{\bar{z}})_{12}(b_z)_{2i} -
(b_z)_{1j}(a_{\bar{z}})_{ji},\ \ \ \ j\geq 3, \nonumber \\
&=& -{\partial}_{\bar{z}}(A_z)_{1i} + (A_{\bar{z}})_{12}(A_z)_{2i} -
(A_z)_{1j}(A_{\bar{z}})_{ji},
\end{eqnarray}
where the structure of (55) and (56) have been used. Using the definition that
$ (A_z)_{1i}\ =\ N^{\mu}_i({\partial}_z{\hat{e}}_1)$ and (16), we find,
\begin{eqnarray}
({\partial}_{\bar{z}}b_z +
[a_{\bar{z}},b_z])_{1i}&=&(A_{\bar{z}})_{12}(A_z)_{2i} +
(A_z)_{12}(A_{\bar{z}})_{2i} -
N^{\mu}_i({\partial}_z{\partial}_{\bar{z}}{\hat{e}}_1).
\end{eqnarray}

   The expressions for $(A_{\bar{z}})_{12}$ , $(A_z)_{2i}$ , $(A_z)_{12}$ and
$(A_{\bar{z}})_{2i}$ have been given in (54). The quantity
$N^{\mu}_i{\partial}_z {\partial}_{\bar{z}}{\hat{e}}_1$ is calculated using
(44) to be,
\begin{eqnarray}
N^{\mu}_i{\partial}_z{\partial}_{\bar{z}}{\hat{e}}_1 &=& \frac{4}{{\mid
\Phi\mid}^4} (O_T)_{ik} [ {\partial}_{\bar{z}}(\frac{1}{\surd 2\mid
\Phi\mid})(A^{\mu}_k{\partial}_z{\Phi}^{\mu} +
A^{\mu}_k{\partial}_z{\bar{\Phi}}^{\mu}) \nonumber \\
&+& {\partial}_z (\frac{1}{\surd 2 \mid \Phi \mid})
(A^{\mu}_k{\partial}_{\bar{z}}{\Phi}^{\mu} +
A^{\mu}_k{\partial}_{\bar{z}}{\bar{\Phi}}^{\mu}) \nonumber \\
&+& (\frac{1}{\surd 2\mid
\Phi\mid})(A^{\mu}_k{\partial}_z{\partial}_{\bar{z}}{\bar{\Phi}}^{\mu} +
A^{\mu}_k {\partial}_z{\partial}_{\bar{z}}{\Phi}^{\mu})].
\end{eqnarray}
Using the expressions for $A^{\mu}_k$ in (47) and ${\Phi}^{\mu}$ in (44), the
above quantity has been evaluated using,
\begin{eqnarray}
A^{\mu}_k{\partial}_z{\Phi}^{\mu} &=& {\mid \Phi \mid }^2{\zeta}_{kz},
\nonumber \\
A^{\mu}_k{\partial}_{\bar{z}}{\Phi}^{\mu} &=& {\mid \Phi \mid
}^2{\zeta}_{k\bar{z}}, \nonumber \\
A^{\mu}_k{\partial}_z{\bar{\Phi }}^{\mu} &=& -4({\bar{\zeta}}_k +
{\zeta}_k{{\bar{\zeta}}_q}^2)({\zeta}_j+{\bar{\zeta}}_j{{\zeta}_m}^2)
{\bar{\zeta}}_{jz} + {\mid \Phi \mid }^2({\bar{\zeta}}_{kz} +
2{\zeta}_k{\bar{\zeta}}_m{\bar{\zeta}}_{mz}) \nonumber \\
A^{\mu}_k{\partial}_{\bar{z}}{\bar{\Phi}}^{\mu} &=& -4({\bar{\zeta}}_k +
{\zeta}_k {\bar{\zeta}}^2_{q})({\zeta}_j +
{\bar{\zeta}}_j{\zeta}^2_{m}){\bar{\zeta}}_{j\bar{z}} + {\mid \Phi \mid
}^2({\bar{\zeta}}_{k\bar{z}} +
2{\zeta}_k{\bar{\zeta}}_m{\bar{\zeta}}_{m\bar{z}}) \nonumber \\
A^{\mu}_k{\partial}_z{\partial}_{\bar{z}}{\Phi}^{\mu} &=& 4({\bar{\zeta}}_k +
{\zeta}_k{\bar{\zeta}}^2_m){\zeta}_{qz}{\zeta}_{q\bar{z}} + {\mid \Phi \mid }^2
{\zeta}_{kz\bar{z}} \nonumber \\
A^{\mu}_k{\partial}_z{\partial}_{\bar{z}}{\bar{\Phi}}^{\mu} &=&
-4({\bar{\zeta}}_k + {\zeta}_k{\bar{\zeta}}^2_m)[({\zeta}_j +
{\bar{\zeta}}_j{\zeta}^2_q){\bar{\zeta}}_{jz\bar{z}} + {\zeta}^2_q
{\bar{\zeta}}_{j\bar{z}}{\bar{\zeta}}_{jz}] \nonumber \\
& + & {\mid \Phi \mid }^2[2{\zeta}_k{\bar{\zeta}}_{j\bar{z}}{\bar{\zeta}}_{jz}
+ 2{\zeta}_k{\bar{\zeta}}_j{\bar{\zeta}}_{jz\bar{z}} +
{\bar{\zeta}}_{kz\bar{z}}],
\end{eqnarray}
and the Euler-Lagrange equations of motion (52) (harmonic map requirement) for
the $z\bar{z}$-derivatives of $\zeta$'s. We then find,
\begin{eqnarray}
({\partial}_{\bar{z}}b_z + [a_{\bar{z}} , b_z])_{1i} &=& 0,
\end{eqnarray}
which is the required self-duality equation. Similarly the other components
have been verified. This proves our main result that for immersed surfaces in
$R^n$ for $n>4$, the surface $M$ admits Hitchin's self-dual system when the
Gauss map is harmonic. Explicit solutions to the self-dual equations for the
gauge group $SO(2)\times SO(n-2)$ are given by (55) and (56), where the complex
functions $\zeta$'s satisfy the equation of motion (52).

\vspace{1.5cm}

\noindent {\bf V.CONCLUSIONS}
\vspace{0.5cm}

   We have considered certain properties of oriented  string world sheet
immersed in $R^{n}$ . The immersion is described by the Gauss map from
the Riemann surface induced by the conformal structure on the immersed
surface into $G_{2,n}$. For $n=3$ and $n=4$, when the Euler-Lagrange equations
following from the extrinsic curvature action expressed in terms of the
coordinates of the  Grassmanian  are satisfied, the Gauss map is harmonic
and the mean curvature scalar of the immersed surface is constant. For such
a class of surfaces, we have made use of the $SO(3)$ and $SO(4)$ two
dimensional gauge fields constructed by us [4] and projected them onto
the subgroup and its orthogonal complement in the Grassmannian  $G_{2,3}$
and  $G_{2,4}$. The projection onto the subgroup transforms as a gauge field
belonging to the subgroup while the complement transforms homogeneously.
By identifying the complement with the complex Higg'sfield, we are able
to prove the existence of solutions to Hitchin's self-dual
equation for  constant $h$ immersions in $R^3$ and $R^4$. This study
complements our earlier result that $h\surd g$ = 1, surfaces exhibit Virasaro
symmetry. The quantum theory of these surfaces have recently been studied
by us [5].

   The self-dual system so obtained for harmonic maps is compared with the
self-dual Chern-Simons system. A generalized Liouville equation involving
extrinsic geometry is obtained. As a particular case, when the map is
anti-holomorphic the familiar Toda equations are obtained.

   We have generalized the results to conformal immersion of 2-dimensional
surfaces in $R^n$, for arbitrary n, using the results of the generalized Gauss
map. We prove
that the surface has constant mean curvature when the Gauss map is harmonic.
This harmonicity condition or Euler-Lagrange equation is used to show that for
such surfaces, there exists Hitchin's self-dual system.

   The general action for the string theory will be a sum of the Nambu-Goto
action and the action involving extrinsic geometry. When the theory is
described in terms of the Gauss map, we have earlier [2] noted that both the
actions can be expressed as Grassmannian sigma model. Explicitly,
\begin{eqnarray}
S &=& S_{NG} + S_{Extrinsic} \nonumber \\
  &=&\mu \int \surd g dz\wedge d\bar{z} +\sigma \int \surd g {\mid H \mid }^2
dz\wedge d\bar{z}, \nonumber \\
  &=&\mu \int \frac{1}{h^2} g_{ij}{\zeta}_{i\bar{z}}{\bar{\zeta}}_{jz} dz\wedge
d\bar{z} + \sigma \int g_{ij}{\zeta}_{i\bar{z}}{\bar{\zeta}}_{jz} dz\wedge
d\bar{z}.
\end{eqnarray}
In general the mean curvature scalar $h$ will be a (real) function of $\zeta$
and so the study of the action $S$ will be complicated since we have a
space-dependent coupling for $S_{NG}$ in this frame work. When surfaces of
constant $h$ are considered or when the Gauss map is harmonic, it is easy to
see that the total action is just a Grassmannian sigma model with one effective
coupling constant. Since the classical equations of motion for this action are
identical to (52), it is possible to study the quantization in the background
field method.

\vspace{0.5cm}

\noindent {\bf Acknowledgements}

We are thankful to Prof.I.Volovich and Prof.I.Y.Arefeva for useful discussions.
One of us (R.P) wishes to thank Prof.C.S.Seshadri for valuable discussions.
This work has been supported by an operating grant (K.S.V) from the Natural
Sciences and Engineering Council of Canada. R.P thanks the Department of
Physics, Simon Fraser University, for the hospitality during the summer 1993.

\vspace{0.5cm}

\noindent {\bf References}

\begin{enumerate}
\item N.J.Hitchin, Proc.London.Math.Soc.(3){\bf 55},59(1987)
\item K.S.Viswanathan,R.Parthasarathy and D.Kay,Ann.Phys.(N.Y){\bf \\
206}237(1991)
\item D.A.Hoffman and R.Osserman,J.Diff.Geom.{\bf 18},733(1983); \\
Proc.London.Math.Soc.(3){\bf 50},21(1985)
\item R.Parthasarathy and K.S.Viswanathan,Int.J.Mod.Phys.{\bf A7},317(1992)
\item K.S.Viswanathan and R.Parthasarathy, Int.J.Mod.Phys.{\bf A}(submitted)
\item S.K.Donaldson,Proc.London.Math.Soc.(3){\bf 55},127(1987)
\item E.A.Ruh and J.Vilms,Trans.Amer.Math.Soc.{\bf 149},569(1979)
\item G.V.Dunne,R.Jackiw,S-Y.Pi and C.A.Trugenberger, Phys.Rev.{\bf D43}, \\
1332(1991)
\item G.V.Dunne, Comm.Math.Phys.{\bf 150},519(1992)
\item K.Kenmotsu, Math.Ann.{\bf 245},89(1979)
\item R.Parthasarathy and K.S.Viswanathan,Int.J.Mod.Phys.{\bf A7},2819(1992)
\item D.A.Hoffman and R.Osserman,Mem.Amer.Math.Soc.No.236 (1980).
\item A.P.Balachandran,A.Stern and G.Trahern,Phys.Rev.{\bf D19},2416(1979)
\item K.Fujii,Lett.Math.Phys.{\bf 25},203(1992)
\end{enumerate}
\end{document}